\documentclass[12pt,preprint]{aastex}
\shorttitle{Cluster Ellipticities}
\shortauthors{Ho, Bahcall \& Bode}

\begin{document}

\title{Cluster Ellipticities as a Cosmological Probe}
\author{Shirley Ho \altaffilmark{1}, Neta Bahcall \& Paul Bode}
\affil{Department of Astrophysical Sciences, 
Princeton University, NJ 08544}
\altaffiltext{1}{E-mail: shirley@astro.princeton.edu}

\begin{abstract}
We investigate the dependence of ellipticities of clusters of galaxies
on cosmological parameters using large-scale cosmological simulations.
We determine cluster ellipticities out to redshift unity
for LCDM models with different mean densities $\Omega_m$ and 
amplitudes of mass fluctuation $\sigma_{8,0}$. 
The mean ellipticity increases monotonically with redshift for all models.
Larger values of $\sigma_{8,0}$, i.e., earlier cluster
formation time, produce lower ellipticities. 
The dependence of ellipticity on $\Omega_m$ is relatively weak 
in the range $0.2 \leq \Omega_m \leq 0.5$ for high mass clusters.
The mean ellipticity  $\bar{e}(z)$ decreases linearly with
the amplitude of fluctuations at the cluster redshift $z$,
nearly independent of $\Omega_m$;
on average, older clusters are more relaxed and are thus less
elliptical.
The distribution of ellipticities about the mean
is approximated by a Gaussian,
allowing a simple characterization of the evolution
of ellipticity with redshift as a function of 
cosmological parameters.
At $z=0$, the mean ellipticity of high mass clusters
is approximated by 
$\bar{e}(z=0) = 0.248-0.069 \sigma_{8,0} + 0.013 \Omega_{m,0}$.
This relation opens up the possibility that,
when compared with future observations of large cluster samples, 
the mean cluster ellipticity
and its evolution  could be used as a new, independent tool to constrain
cosmological parameters,  especially the amplitude of mass
fluctuations, $\sigma_{8,0}$.
\end{abstract}

\keywords{cosmology: theory --- galaxies: clusters: general ---
large-scale
structure of the universe}

\section{Introduction}
Over the last twenty years the Cold Dark Matter (CDM) paradigm
has become the standard model for structure formation in the
Universe. It assumes the cosmic mass budget to be dominated by
CDM, whose gravitational effects build structure
from an initially Gaussian distribution of adiabatic fluctuations.
Important parameters specifying CDM models include
the fraction of the critical density in matter, $\Omega_m$, and
the fraction of the critical density in dark energy, $\Lambda$;
these in part determine the expansion rate of
the universe and the shape of the initial matter power spectrum.
Another important parameter is
the amplitude of the power spectrum, conventionally quoted in 
terms of the rms linear mass fluctuation at $z=0$ in a sphere of 
radius 8 $h^{-1}$Mpc,  $\sigma_{8,0}$
($H_0=100h$\,km\,s$^{-1}$Mpc$^{-1}$ throughout).

A series of exciting observations have been made in recent years
which constrain these parameters,
resulting in a concordance model \citep{BOPS99,Spergel03} ---
a spatially flat $\Lambda$CDM model with $\Omega_m\sim0.25$.
These include 
measurements of the cluster mass function, the strong cluster correlation 
function, and the evolution of the cluster abundance with redshift
\citep{frenk90,bahcall92,eke98,viana02,SDSSmf03,BB03};
the magnitude-redshift relation
of Type Ia supernovae \citep{Riess,Perlm};
optical surveys of large scale structure 
\citep{Eisenstein05,Tegmark,PopeSDSS04,Cole2df05};
anisotropies in the comic microwave
background
\citep{bennett96,Spergel03};
cosmic shear from weak lensing observations 
\citep{vWaerb01,Refregier03,vWM05};
and Lyman-$\alpha$ forest absorption
\citep{croft99,mcdonald04}.

A key concept in the build-up of structure
in this model is the formation of
dark matter halos---  quasi-equilibrium systems of dark
matter, formed through non-linear gravitational collapse.
Galaxies and other luminous objects are assumed to form by cooling
and condensation of baryons within these halos 
\citep{wr78}.
Thus understanding the evolution and development of dark matter halos is
an important step toward understanding structure formation.
The possibility of using internal cluster properties 
to place constraints on cosmological parameters has
been explored in a variety of ways. For example \citet{RLT92} 
suggested that the degree of substructure in clusters
could constrain $\Omega_m$.  Different measures 
of substructure have been examined by \citet{CER96}
and \citet{BuoTsa95}; the power ratio method of the
latter has been applied to Chandra data by \citet{JCBB05},
showing that clusters had more substructure in the past,
but without setting any cosmological constraints.

An interesting property to investigate 
is the ellipticity or axis ratio of clusters.
Observations
suggest that cluster shapes, as traced by
the distribution of galaxies \citep{PBF91,RvHK91,WestBoth90,SPPDDeCG05}, 
X-ray emission and/or SZ decrement 
\citep{MKU89,MEFG95,KBPG01,WangFan04,DeFSBL05,FAKPBB05}, 
and gravitational lensing \citep{OLS03,henk04,mandel05}, 
are usually not spherical.  There is some
indication that the mean
ellipticity is evolving with redshift \citep{melott01,plionis02}.
Dark matter and hydrodynamic
simulations have also been utilized, 
resulting in cluster-sized halos which
are triaxial 
\citep{WDO89,EMFG93,dTKvK95,SMLBT97,BuoXu97,JS02,floor03,suwa03,hw04,
rsmm04,FAKPBB05,HBB05,KasEvr05,AFPKWFB05}.

In this paper, we investigate the possibility
of using cluster ellipticities as a cosmological probe,
in particular of $\Omega_m$ and $\sigma_{8,0}$.
The observed cluster samples are currently too small and too poorly
characterized for strong constraints to be developed.
However, with the ongoing optical, X-ray, weak lensing and SZ surveys 
providing an ever-increasing sample of clusters
\citep[e.g.][]{Betal03,MillerSDSS05,Breflex04,CHR02,SPTref,APEX,ACT03}
it will become possible to make more accurate
statistical inferences concerning cluster
ellipticities.
In this paper we provide predictions of cluster ellipticities 
and their evolution for different cosmologies that could be 
directly compared with observations. 
We use numerical simulations of structure formation to 
generate clusters and
determine their ellipticity evolution with redshift from $z=0$ to
$z=1$,
systematically varying $\Omega_m$ and $\sigma_{8,0}$ in
order to understand how these parameters affect cluster structure.
The simulations, cluster selection,
and derivation of cluster ellipticity are presented in
\S\ref{sec:ellip}.
The results of cluster ellipticities are  shown in \S\ref{sec:resellip},
and we conclude in \S\ref{sec:conc}.

\section{Predicted cluster ellipticities} \label{sec:ellip}

A series of N-body LCDM cosmological simulations were run, with the only
differences between runs being in the matter density and
the power spectrum amplitude;  these were set to
$\Omega_m=$ 0.2, 0.3, and 0.5, with $\sigma_{8,0}=$ 0.7, 0.9, and 1.1,
for nine models in all.  The linear CDM power spectrum
was
generated using the lingers.f code from the
GRAFIC2\footnote{Available at \url{http://arcturus.mit.edu/grafic/}}
package \citep{Bert01}, with a Hubble constant
$h=0.7$, baryon density $\Omega_b=0.041$,
and spectral index $n=1$.
GRAFIC2 was then used to generate the initial particle conditions,
with the modification that the Hanning filter was not used
because it suppresses power on small scales \citep{MTC05}.
All of the runs contained $N=256^3$ particles in a periodic
cube of size $500h^{-1}$Mpc, making the particle mass
$m_p=2.07\times10^{12} \Omega_m h^{-1}M_\odot$.
Simulations were carried out using the
TPM\footnote{Available at
\url{http://astro.princeton.edu/$\sim$bode/TPM/}}
code \citep{BO03} with a $512^3$ mesh and a spline softening
length of 20.35$h^{-1}$kpc.  The initial domain decomposition
parameters in the TPM code were $A=1.9$ and $B=8.0$ (TPM was
modified slightly so that there was no lower limit to $B$ 
when it is reduced at later times, which improves the
tracking of low mass halos; 
for details on these parameters see \citet{BO03}).

The particle positions were saved for every expansion
factor interval 0.1 between $a=0.5$ and $a=1$, i.e. from
$z=1$ to the present.
The FOF (friends-of-friends) halo finder was run on
each saved redshift, with a linking length
$b=0.164$ times the mean interparticle separation
\citep{LC94,JFWCCEY01}.
The resulting mass functions were in good agreement
with the predictive formula of \citet{JFWCCEY01}
(using $b=0.164$) for halos with 32 or more particles
(or mass above $3.3\times10^{13}h^{-1}M_\odot$ for
$\Omega_m=0.5$).  The center of mass of each FOF halo
thus identified was calculated, and all particles in
a cube of size 4$h^{-1}$Mpc around this point were
extracted.  The most bound particle in this cube
was identified, and all the particles within 1$h^{-1}$Mpc
of this particle were selected to compute the ellipticity.
All cluster ellipticities in this paper will refer to the ellipticity
within a comoving radius of $1h^{-1}$Mpc,
as this radius should be easier to determine observationally.
Only halos with mass
above $4\times10^{13}h^{-1}M_\odot$
(39 particles for $\Omega_m=0.5$) within the 1$h^{-1}$Mpc radius
were considered.
Varying $b$ from 0.14 to 0.2 had little effect on
the final masses and ellipticities, because once a
cluster center is chosen we use the mass within a
specific radius of 1$h^{-1}$Mpc.
The same set of random phases was used for all the runs;
we experimented with varying the seed used to generate these
phases, and found this had no impact on the results.
Current observations indicate that the dark matter radial 
profile of clusters, obtained by gravitational lensing
observations, follows the galaxy radial profile
on the scale of 1$h^{-1}$Mpc \citep{fis97,CYE97}.
It is therefore expected that the observed distribution
of galaxies will closely reflect the underlying mass distribution.
The underlying mass distribution is also reflected in the 
observed distribution of the intracluster gas 
\citep[for a recent review see][]{Arnaud2005}.
Further observations will shed additional light 
on this comparison.

To determine the ellipticity of a cluster,
we find the best-fit ellipse using
the matrix of second moments of particle positions
about the center of mass:
\begin{equation}
I_{ij} = \sum m_p x_i x_j
\end{equation}
where the sum is over all the particles selected in the manner
just described.
Given the normalized eigenvalues of $I$, $\lambda_i$, we use
a common measure of ellipticity:
\begin{equation}
\epsilon= 1- \sqrt {\lambda_2/ \lambda_1}  = 1-a_2/a_1 
\end{equation}
where $\lambda_1>\lambda_2>\lambda_3$, and $a_1$ and $a_2$ are the 
primary and secondary ellipsoid axes' lengths respectively.
Since
the 3-D ellipticity is not directly measured in observations, 
we also measure the projected 2-D cluster ellipticities  
on the x-y plane.
The process of determining the projected cluster ellipticities is
repeated as above and the 2-D ellipticity of each halo is:
\begin{equation}
\epsilon_{2d}= 1- \sqrt {\lambda_{2d,2}/ \lambda_{2d,1}}
\end{equation}
where $\lambda_{2d,2} > \lambda_{2d,1}$.
This 2-D ellipticity will be our definition of ellipticities in the paper, unless
otherwise specified.

After selecting the clusters as described above,
we further separate the clusters into two groups:
low mass clusters with masses $M_{1.0}$ in the range
$4\times 10^{13}\leq M_{1.0}<10^{14}h^{-1}M_{\odot}$,
and high mass clusters with $M_{1.0}\geq 10^{14}h^{-1}M_{\odot}$
($M_{1.0}$ denotes the mass within $1 h^{-1}$Mpc).
This division was made because
the mean cluster ellipticities have a small mass dependence.
Moreover, cluster surveys (including ACT, APEX, and SPT)
which will produce
larger cluster samples, will be best at detecting clusters
with masses above
$10^{14}$ $h^{-1}M_{\odot}$ \citep{SPTref}.
As it would be challenging to have accurate measures of mass
for a large number of clusters, we have investigated the effect of 
a $30\%$ uncertainty in the mass
determination in the threshold cut of $10^{14}$ $h^{-1}M_{\odot}$;
this change affected the resulting
mean ellipticities by less than 3\%.

\section{Cluster Ellipticities and Structure Growth} \label{sec:resellip}

\begin{figure}
\plotone{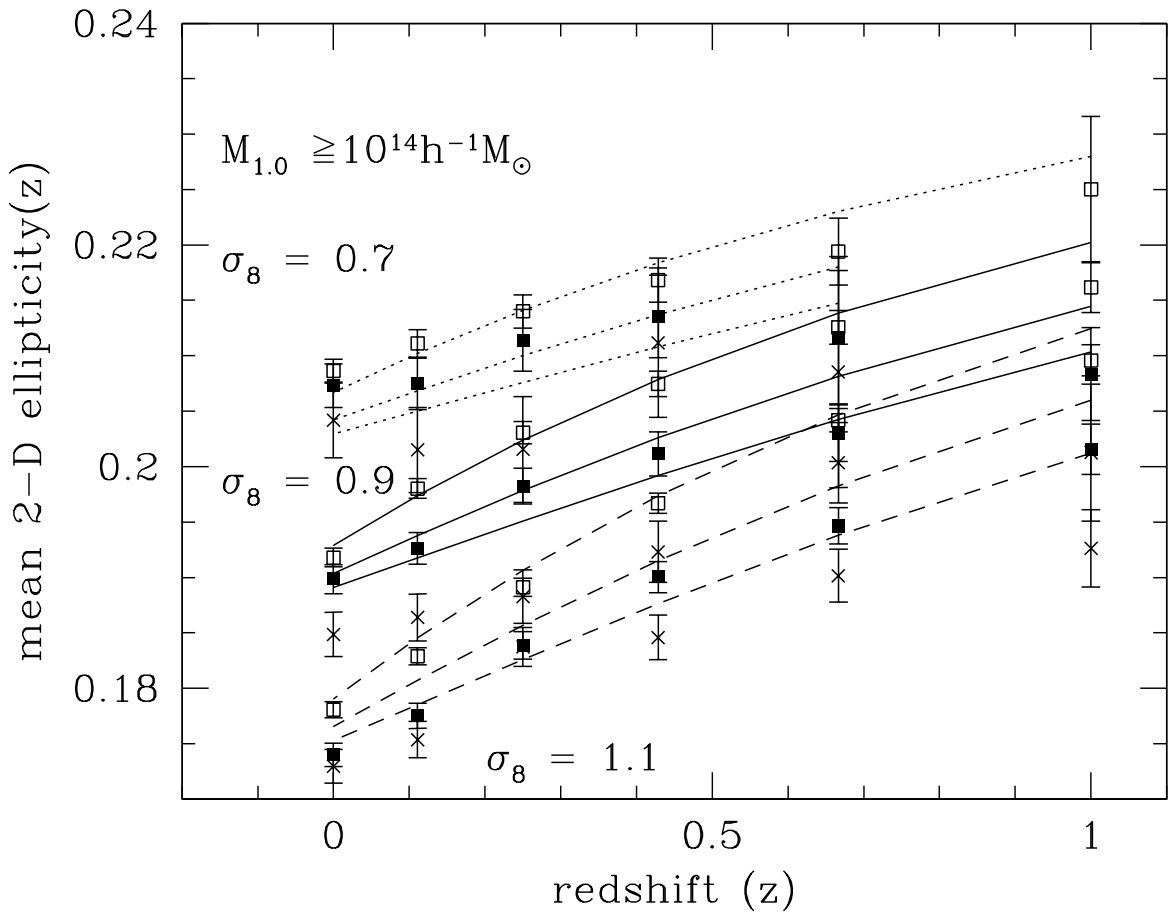}
\caption{The evolution of mean cluster ellipticity
(for $M_{1.0}\geq 10^{14} h^{-1}M_{\odot}$ clusters) as a function
of redshift, for different models.
The crosses represent 
$\Omega_{m}= 0.2$;
solid squares represent $\Omega_{m}=0.3$;
empty squares $\Omega_{m}= 0.5$.
The lines represent the corresponding
fits from Eqn. \ref{eqn:hmfit}:  
short-dash lines for $\sigma_{8,0} = 0.7$,
solid lines for $\sigma_{8,0}=0.9$,
long-dash lines for $\sigma_{8,0} = 1.1$.
Error bars are the variance of the mean. }
\label{fig:E_z_hm}
\end{figure}

In this section we investigate how cluster ellipticities 
depend  upon the cosmological parameters $\Omega_m$ and $\sigma_{8,0}$.   
Fig. \ref{fig:E_z_hm} presents the mean ellipticities as a 
function of redshift for high mass
clusters ($M_{1.0}\geq 10^{14}h^{-1}M_{\odot}$).
The error bars show the variance of the mean:
$\sigma_e = N^{-1}\sqrt{\sum_i^N (e_i-\bar{e}(z))^2}$.
The mean ellipticity
increases with redshift for all models considered here.
The dependence of cluster ellipticities on $\sigma_{8,0}$ is significant:
higher $\sigma_{8,0}$ values (i.e. clusters are forming earlier)
lead to lower ellipticities at $z=0$.
In contrast, the dependence on $\Omega_m$ is relatively weak. 

\begin{figure}
\plotone{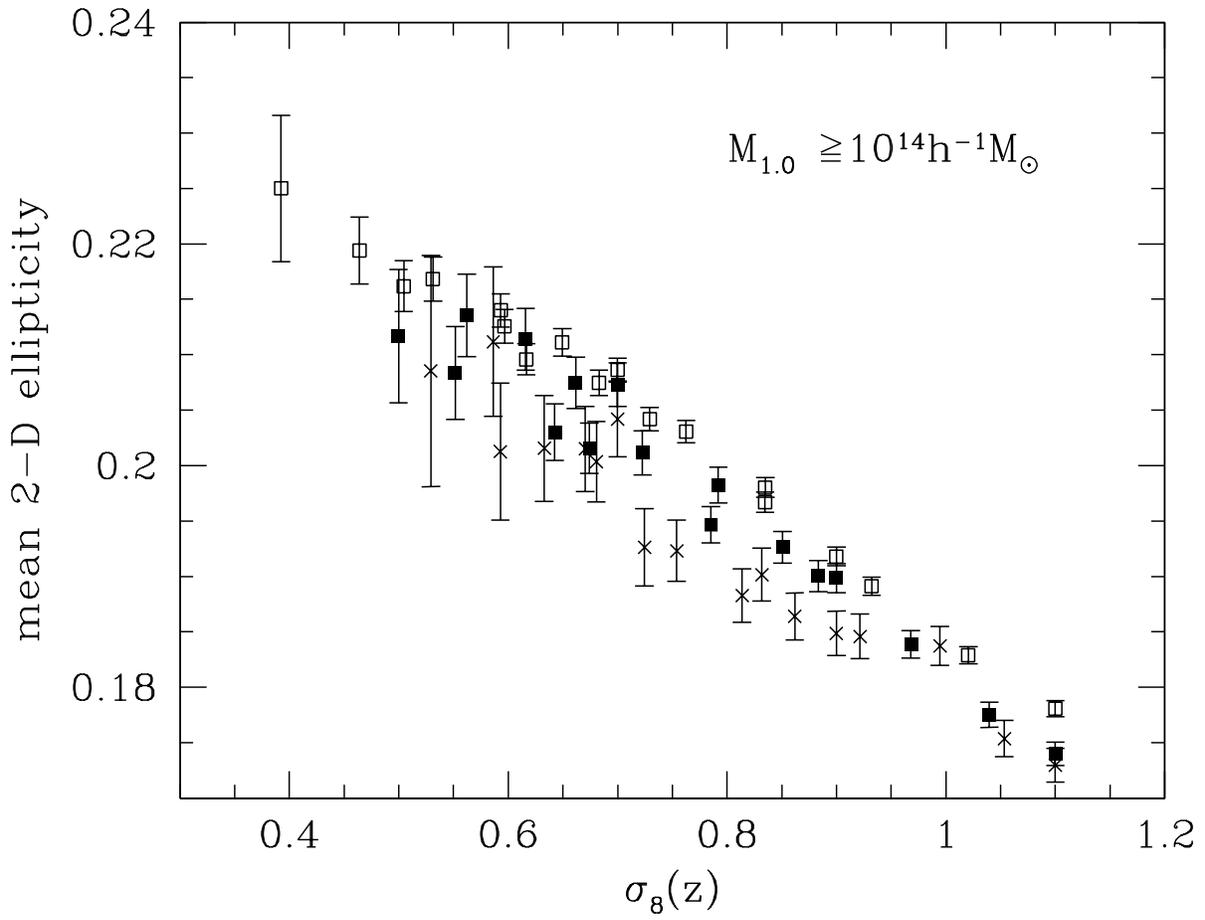}
\caption{The mean cluster ellipticity $\bar{e}(z)$
at various redshifts $z\leq 1$,
as a function of the linear $\sigma_8(z)$ 
(amplitude of mass fluctuations at redshift $z$).
Point types denote different $\Omega_m$,
as in Fig. \ref{fig:E_z_hm}; all nine models are shown.
Only clusters with
$M_{1.0}\geq 10^{14} h^{-1}M_{\odot}$ are included.
}
\label{fig:E_s8_hm}
\end{figure}

An intriguing result is apparent in Fig. \ref{fig:E_s8_hm},
which  plots mean ellipticity at $z$,
$\bar{e}(z)$, as a function of
$\sigma_8(z)$: there is an inverse linear relation
between the two, nearly independent of $\Omega_m$
(except indirectly in that $\sigma_8(z)$ depends on $\Omega_m$).
While the formation of clusters is a complex process,
it appears that
some average properties of clusters, such as ellipticities,
can still be understood with a fairly simple picture:
on average, the older the cluster, the more relaxed it is, and thus 
less elliptical.
This is of course an over-simplified picture of the formation 
of clusters, but it does capture the general relation of 
Fig.  \ref{fig:E_s8_hm}.
Based on this Figure,
we fit the mean ellipticity $\bar{e}(z)$ of the high mass clusters
($M_{1.0} \geq 10^{14} h^{-1}M_{\odot}$)
to the following
linear relationship:
\begin{equation} \label{eqn:hmfit}
\bar{e}(z) = 0.248\left( 1 - 0.250\frac{\sigma_8(z)}{0.9}
+ 0.016\frac{\Omega_m}{0.3}
\right)
\end{equation}
At $z=0$ this relation becomes 
$\bar{e}(z=0) = 0.248-0.069 \sigma_{8,0} + 0.013 \Omega_{m,0}$.
This relation of  ellipticity as a function
of redshift is shown by the lines
in Fig. \ref{fig:E_z_hm}.  While there are
small variations from model to model, the general trends
are followed.

The distribution of ellipticities can be fit reasonably
well by a Gaussian for all $z\leq 1$.  For high mass
clusters, the standard deviation is nearly the same for
all runs:  a value of $\sigma=0.09$ is within ten percent
of the value measured at any redshift for all the models.
Thus, using the mean from Eqn. \ref{eqn:hmfit}, it is
possible to characterize the complete
distribution of ellipticities out to $z=1$.
Also, as mentioned earlier, we have investigated including
a $30\%$ uncertainty in the mass
determination in the cut at $10^{14}$ $h^{-1}M_{\odot}$;
this did not affect the resulting
mean ellipticities at more than 3\%.

\begin{figure}
\plotone{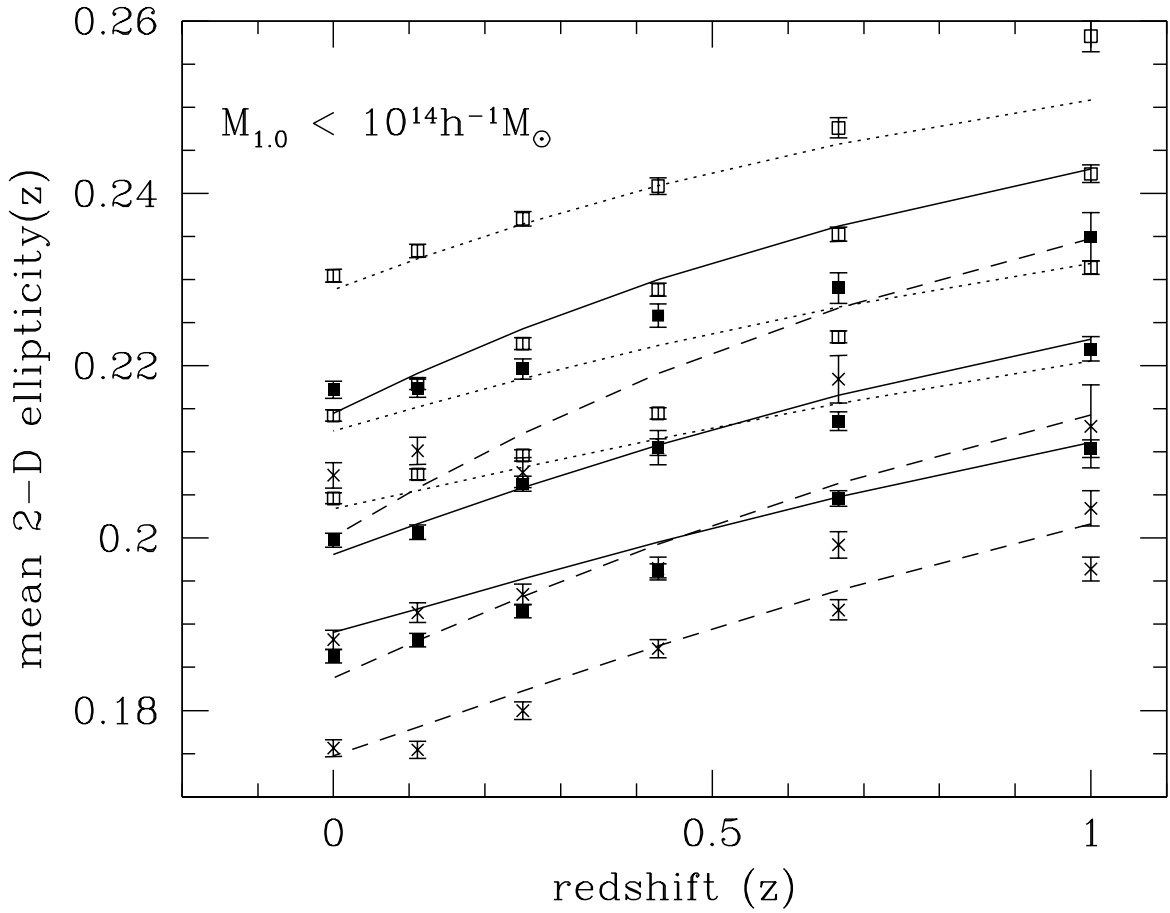}
\caption{The evolution of mean cluster ellipticities,
for low mass clusters.  Point and line types as in Fig. \ref{fig:E_z_hm}
}
\label{fig:E_z_lm}
\end{figure}
\begin{figure}
\plotone{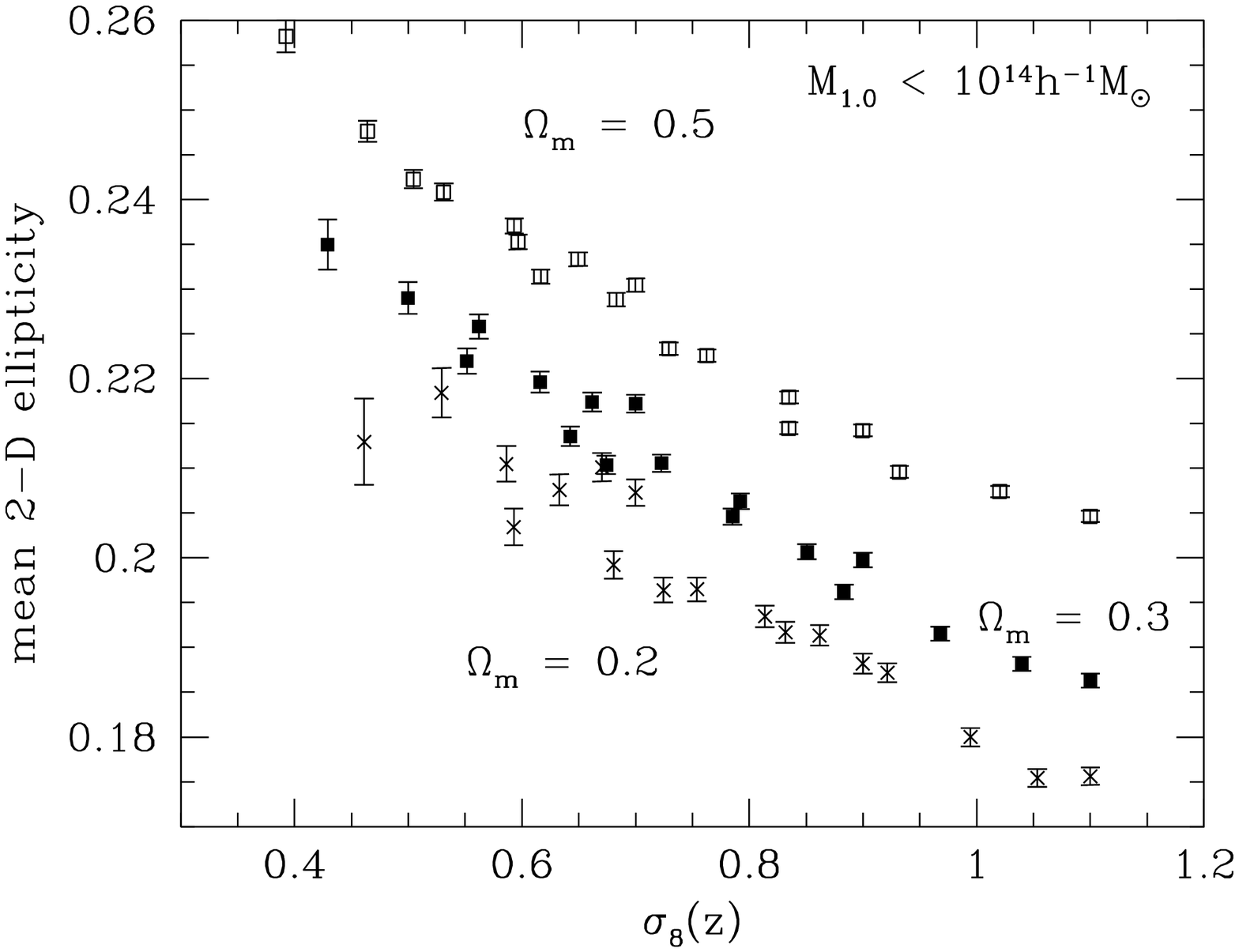}
\caption{The evolution of $\bar{e}(z)$ for
$4\times 10^{13}h^{-1}M_{\odot}\leq M_{1.0}<10^{14}h^{-1}M_{\odot}$,
as a function of $\sigma_{8,0}$; point types as in  Fig. \ref{fig:E_z_hm}. }
\label{fig:E_s8_lm}
\end{figure}

The results for cluster ellipticities of low mass clusters 
($4\times 10^{13}\leq M_{1.0}<10^{14}h^{-1}M_{\odot}$)
are presented in  
Figures \ref{fig:E_z_lm} and \ref{fig:E_s8_lm}.
These Figures demonstrate  that the low mass clusters 
show the same trend of increasing ellipticity with redshift, 
as well as increasing ellipticity with decreasing $\sigma_{8,0}$. However
the dependence on $\Omega_m$ is stronger 
than for the higher mass clusters.
The fit corresponding to Eqn. \ref{eqn:hmfit} for these clusters is:
\begin{equation} \label{eqn:lmfit}
\bar{e}(z) = 0.238\left( 1 - 0.273\frac{\sigma_8(z)}{0.9} 
+ 0.107\frac{\Omega_m}{0.3}
\right)
\end{equation}
While the dependence on $\Omega_m$ is more important for these smaller clusters,
the dependence on $\sigma_8(z)$ is nearly the same for both samples:
as with the higher mass clusters,
for a given $\Omega_m$, the ellipticity decreases as 
$\sigma_8(z)$ increases.

Different definitions of the mass of the cluster
and different methods of measuring the axis ratio
will change the measured value of the mean ellipticity.
It has generally been found in simulations that less
massive objects are rounder \citep[e.g.][]{JS02,HBB05,AFPKWFB05}.
However, this assumes that the ellipticity is measured using
a fixed overdensity or fraction of the virial radius;  the
outermost radius used thus becomes larger for more massive
halos.  Here we instead employ a fixed annular radius 
of 1$h^{-1}$Mpc, which is more suitable for direct comparison with 
observations. This radius is less than the virial radius for
the largest clusters, while extending beyond the virial radius
for less massive systems.
Thus the low-mass sample,
which includes more of the unrelaxed outlying regions of the clusters,
actually has the same or 
a slightly higher mean ellipticity than the high-mass sample,
where the relaxed cores dominate.
Interestingly, \citet{LJS05} derived an analytic prediction
that more massive halos are less elliptical, but concluded
there were conceptual difficulties in comparing their work
to simulations, concerning precisely
this issue of how the axis ratios are to be determined.
It is of course important when comparing observations to
simulations that the same defintions and methods be 
followed in the comparison.

\section{Discussion and Conclusion} \label{sec:conc}

We use large-scale cosmological simulations to determine the
mean ellipticity of
clusters of galaxies and its evolution with redshift to $z=1$,
in order to investigate the possible
use of this quantity as a new tool in constraining cosmological
parameters.
Nine LCDM cosmological models are studied, with $\Omega_m$ ranging from
$0.2$ to $0.5$, and $\sigma_{8,0}$ ranging from $0.7$ to $1.1$.  
We provide predictions
that can be directly compared with future observations of cluster
ellipticities
and can be used to place new independent constraints on cosmological
parameters.

We find that the mean cluster ellipticity increases monotonically
with redshift for all models. Clusters were more elliptical at 
earlier times.  The mean ellipticity of high mass 
clusters  ($M_{1.0}\geq 10^{14} h^{-1}M_{\odot}$) 
depends most strongly on $\sigma_{8,0}$:  higher values of
$\sigma_{8,0}$, i.e., earlier cluster formation  times,  produce lower
ellipticities than lower $\sigma_{8,0}$ values.  The dependence of
ellipticity on $\Omega_m$ is relatively weak.
The high mass clusters exhibit an interestingly
regular behavior for all the
LCDM models:  the mean ellipticity  $\bar{e}(z)$ depends linearly 
on $\sigma_8(z)$,
the amplitude of fluctuations at the cluster redshift $z$.  The effect
of $\Omega_m$
is weak (in the range $0.2 \leq \Omega_m \leq 0.5$) 
for a given $\sigma_8(z)$.  The
distribution of ellipticities for these clusters can be described by
a Gaussian, with a mean given
by Eqn. \ref{eqn:hmfit} and a standard deviation of 0.09. This simple
description can be
used for direct comparison with observations, as well as for
constraining the cosmological parameters, especially $\sigma_8$. It can
also be a useful input
into the 'halo model' \citep{SW05} of large-scale
structure.

The close connection between the mean ellipticity and
$\sigma_8(z)$, suggests
that clusters form with similar high ellipticities at formation time
and thereafter
undergo relaxation, becoming more spherical with time  e.g.,
\citep[e.g.][]{floor03,HBB05,AFPKWFB05}. The amplitude $\sigma_8(z)$
reflects the
cluster formation time: larger amplitudes correspond to earlier
cluster formation,
and hence to a lower relative ellipticity at a given redshift. In a
low density
universe, furthermore, the merger rate is reduced at low redshift,
allowing this relaxation to continue without much perturbation. Both
the formation time of clusters and the merger rate are linked with
the amplitude of the spectrum of mass fluctuations and hence with the
ellipticity.
However, the exact time of cluster formation cannot be easily defined
\citep{cohn05}, since a more
accurate theoretical  understanding of structure formation is
necessary to explain the link.

Numerous large cluster surveys are currently underway or in planning
stages,
aimed at producing increasingly more accurate and complete
samples of clusters. 
Such  surveys will enable the determination of cluster ellipticities
for
large and complete cluster samples using the galaxy distribution
(optical),
the hot gas distribution (X-ray and SZ), and the dark matter
distribution
(lensing) in the clusters.   These surveys include, among others,
the Sloan Digital Sky Survey \citep{Betal03,MillerSDSS05},
the Red Sequence Cluster Survey \citep{RCSref},
X-ray and SZ cluster surveys \citep{FAKPBB05,SPTref,APEX}, 
ACT \citep{ACT03},
gravitational lensing surveys \citep{Panstars}, 
and LSST\footnote{\url{http://www.lsst.org}}.
Each method has different selection functions
and selection biases, as is well known. The ability to measure and compare
cluster ellipticities using several independent methods will enable not
only an improved understanding of the internal structure and physical
processes in clusters, but will also enable the use of cluster
ellipticities in cosmology.

\acknowledgments
The computations were run at Princeton on facilities supported 
by NSF grant AST-0216105,
and at the National Center for Supercomputing Applications 
with support of a grant of supercomputing time (number MCA04N002P).
This research was supported in part by Bahcall's NSF grant AST-0407305.
We would like to thank Neal Dalal, Hy Trac, Joe Hennawi,
Chris Hirata, Amol Upadhye, Feng Dong,
and Mike Gladders for their helpful discussions.

\end{document}